\documentstyle[epsfig]{aipproc}
\begin{document}
\title{VHE Astronomy before the New Millennium}
\author{Trevor C. Weekes}
\address{Whipple Observatory, Harvard-Smithsonian Center for
Astrophysics, \\P.O. Box 97, Amado, AZ 85645-0097 U.S.A.\\
e-mail: tweekes@cfa.harvard.edu}
\maketitle

\section{Introduction}

In planning this workshop the organizers suggested that a review of
new TeV $\gamma$-ray observations at the beginning 
might obviate a need for separate submissions from individual
groups (which would be presented anyway at the ICRC) and thus
permit the program to be devoted to technical developments,
interpretations and new programs. In practice, this did not work
too well since many groups wished to personally present their own
results at the workshop also. Therefore I tried to present some
synthesis of the results as seen in August, 1999. In principle,
these are pre-workshop but in practice since many of the ICRC
papers were available (courtesy of astro-ph) and others were sent
directly to me, I have been able to include many of the most recent
results. As usual, the printed ICRC papers do not contain the whole
story since many authors still regard the pre-conference papers as
place holders with the real results to be presented orally at the
conference. Hence the proceedings are archaic as soon as they are
printed (or recorded on CD)!

I will concentrate on the TeV observations reported since the last
workshop (Kruger Park, August, 1997). I will not discuss the
impressive technical improvements reported from many 
Very High Energy (VHE)
observatories, nor describe the new VHE
observatories that have come on-line (or are due shortly to come
on-line) nor the many interpretations now reported for the VHE
observations. Suffice to say that VHE astronomy is still
observation-driven and that theoretical VHE astrophysics still lags
as a predictive discipline.

More general reviews of VHE $\gamma$-ray astronomy can be
found elsewhere \cite{ong98,cataneseweekes99}.

\section{Standards of credibility}

It is well known that VHE $\gamma$-ray astronomy has had a rather
murky past in that many sources were claimed in the eighties which
were never verified. In fairness it should be noted that in the
balloon era of 100 MeV $\gamma$-ray astronomy (in the sixties)
there were also many unverified claims which were only resolved
with the advent of satellite astronomy and the use of reliable
statistical methods. Nevertheless VHE $\gamma$-ray astronomy
suffers from the problems that the detector medium is the
atmosphere over which the experimenter has no control and that non-
statistical effects due to atmospheric/detector conditions are a
perpetual problem. This is primarily so for optical
techniques but it has yet to be demonstrated that air shower arrays
are free of such effects. In such circumstances, some subjective
bias in the data analysis, i.e., in data selection, is almost
inevitable. It should also be remembered that this is a relatively
new discipline (it is only ten years since the publication of the
first verifiable observation) and hence its exponents are still
feeling their way. In these circumstances it behooves all involved
to adopt a conservative attitude to claims for the detection of new
sources. 

It was at the ICRC in Hobart, Australia in 1971 that Prof. A.E.
Chudakov suggested that the acceptable standard of credibility for
a new VHE $\gamma$-ray source should be 5 $\sigma$. I have not been
able to find a reference for this statement and I was not at the
conference but I remember well the dismay with which we greeted
this news on the return of our senior colleagues who attended. At
that stage we had managed to accumulate a 3 $\sigma$ result on the
Crab Nebula \cite{fazio73} which we thought bordered on the edge of
credibility. A 5 $\sigma$ result seemed an impossible standard. 

Years later I met Prof. Chudakov at the 14th Texas Conference on
Relativistic Astrophysics; I reminded him of his 5 $\sigma$
credibility criterion (which had not been met on another source in
the interval) and proudly announced that with the atmospheric
Cherenkov imaging technique we now had a 9 $\sigma$ signal from the
Crab. He paused for a moment and then said drolly: "I think I
should have said 10 $\sigma$". He was right of course; a 5 $\sigma$
detection is not always entirely credible. But he was wrong also,
for a 10 $\sigma$ detection by a single experiment using a new
technique is almost as liable to be a systematic effect as a 5
$\sigma$ detection.

To be convincing, I suggest that we adopt a standard where a 
{\it really}
credible source requires a 5 $\sigma$ detection coupled with an
equally significant verification by another experiment. Ideally the
"$\sigma$" should include the best estimate of potential systematic
effects. This would qualify as a Grade A detection. A Grade B
detection would then be the same but without the independent
verification. Grade C would be a strong detection but with some
qualifications: time variability or some other factors which
introduce extra degrees of freedom.

\section{The 1997 TeV Source Catalog}

In his comprehensive rapporteur talk at the 25th ICRC, Chaman Bhat
listed eleven sources of TeV gamma-rays based on observations
reported at that meeting \cite{bhat97}. These were the SNR
remnants, the Crab Nebula, PRS 1706-44, Vela, and SN1006; the AGNs,
Markarian 421, Markarian 501 and 1ES2344+514; the binary, Centaurus
X-3; and the pulsars, PSR B 1259-63, PSR B 1509-58, and PSR 1105-
61. As we shall see below, all except the three pulsars (about
which there have been no further reports of confirmed detections)
appear in the 1999 catalog.

\section{Source Summary}

\subsection{Galactic Sources}
\noindent
{\bf The Crab Nebula}
\vspace*{0.2cm}

There is remarkable agreement now between the absolute fluxes and
spectral shapes reported from observations of the Crab Nebula by
imaging ACTs; the results from the Whipple, HEGRA, CAT and CANGAROO
experiments are shown in Table \ref{crab-table}. These are also in
agreement with the flux reported in the first detections of the
Crab \cite{weekes89,vacanti91} but this must be considered
fortuitous in view of the large error bars in these early
measurements. At least in the 300 GeV to 3 TeV range it is clear
that the Crab can now be considered a standard candle and a grade  
A source.
  
New observations of the Crab Nebula have been reported at both high
and low energies. CELESTE, with a threshold energy of 50 GeV,
observed it for just three hours \cite{musquere99b}
whereas STACEE, with an interim
threshold of 75 GeV, had a 7 $\sigma$ detection in 50 hours of
observation \cite{oser99}. Neither experiment could quote a flux value and
neither experiment saw any evidence for a pulsed component from the
Crab pulsar.

At higher energies the Crab was seen for the first time by a
conventional air shower array (the Tibet High Density Array at 4.5
km) \cite{amenomori99}. The energy threshold was 3 TeV and the flux
deduced (see Table \ref{crab-table}) was a factor of 2-3 higher
than that seen in ACT experiments.

\begin{table}
\caption{Flux from the Crab Nebula}
\label{crab-table}
\begin{tabular}[t]{lll} 
Group & VHE Spectrum
 & E$_{\rm th}$ \\ 
 & ($10^{-11}$ photons cm$^{-2}$ s$^{-1}$ TeV$^{-1}$) & 
(TeV) \\ 
\tableline
Whipple (1991)\cite{vacanti91}
 & ${\rm   (25 (E/0.4TeV))^{-2.4 \pm 0.3}}$
 & 0.4 \\
Whipple (1998)\cite{hillas98} & ${\rm  
  (3.2 \pm 0.7) (E/TeV)^{(-2.49 \pm 0.06_{stat} \pm 0.04_{syst}}}$
& 0.3 \\
HEGRA (1999)\cite{konopelko99a}  & ${\rm 
  (2.7 \pm 0.2 \pm 0.8) (E/TeV)^{-2.60 \pm 0.05_{stat} \pm
0.05_{syst}}}$ 
 & 0.5 \\
CAT (1999)\cite{musquere99} & ${\rm 
 (2.7 \pm 0.17 \pm 0.40) (E/TeV)^{-2.57 \pm 0.14_{stat} \pm
0.08_{syst}}}$
 & 0.25 \\
CANGAROO (1998)\cite{tanimori98b} & ${\rm
(2.01 \pm 0.36)x10^{-2}) (E/7TeV)^{-2.53 \pm 0.18}}$
& 7 \\
Tibet HD (1999)\cite{amenomori99} & ${\rm
(4.61 \pm 0.90)x10^{-1} (E/3TeV)^{-2.62 \pm 0.17}}$ 
& 3 \\
\end{tabular}
\end{table}

\vspace*{0.2cm}
\noindent
{\bf PSR 1706-44}
\vspace*{0.2cm}

Following the TeV detection of this source by the CANGAROO group
\cite{kifune95} and its confirmation by the Durham group
\cite{chadwick97}, there have been no new reports of observations.
No periodic emission is seen and it is believed
that the VHE emission comes from a weak plerion. Although weaker
than the Crab this may be the standard candle for the southern
hemisphere and merits a grade A ranking.

\vspace*{0.2cm}
\noindent
{\bf SN1006}
\vspace*{0.2cm}

In 1997 the CANGAROO Collaboration reported the observation of TeV
$\gamma$-ray emission from the shell-type SNR, SN\,1006
\cite{tanimori98a}.  Observations taken in 1996 and 1997 indicated
a statistically significant excess from the northeast rim of the
SNR shell. The flux at $> 1.7 \pm 0.5$ TeV was ($4.6  \pm 0.6 (sys)
\pm 1.4 (stat) \times 10 ^{-12}$ photons cm$^{-2}$ s$^{-1}$. The
observations were motivated by the observation of non-thermal X-
rays by the {\it ASCA} experiment \cite{koyama95}.  It represented
the first direct evidence
of acceleration of particles to TeV energies in the shocks of SNRs.

At this workshop there was the disturbing report from the Durham
group of the failure to detect this source in 40 hours of
observation. Their upper limit at 300 GeV was 1.7 $ \times 10 ^{-
12}$ photons cm$^{-2}$s$^{-1}$ and at 1.5 TeV was 1.3 $ \times 10
^{-12}$ photons cm$^{-2}$s$^{-1}$, barely compatible with the
CANGAROO observation. They point out that the presence of a bright
star near the SNR complicates the measurement. Because of this
report I assign this source a B$-$ grade.

\vspace*{0.2cm}
\noindent
{\bf Vela}
\vspace{0.2cm}

The CANGAROO group reported the detection of a 6$\sigma$
signal from the vicinity of the Vela pulsar \cite{yoshikoshi97}.
The
integral $\gamma$-ray flux above 2.5 TeV is $2.5 \times 10^{-12}$
photons cm$^{-2}$ s$^{-1}$. There is no evidence for periodicity
and the flux limit is about a factor of ten less than the steady
flux. The signal is offset (by 0.14$^{\circ}$) from the pulsar
position which makes it more likely that the source is a
synchrotron
nebula. Since this offset position is coincident with the
birthplace
of the pulsar it is suggested that the progenitor electrons are
relics
of the initial supernova explosion and they have survived because
the
magnetic field was weak.

Again the source was not confirmed by observations by the Durham
group (J.Osborne, this workshop). The
upper limit to the $\gamma$-ray flux above 300 GeV is $5 \times
10^{-11}$ photons cm$^{-2}$ s$^{-1}$. Given the differences in
energy and the uncertainties in flux estimates in the two
experiments, the Durham group felt the two results were compatible.
However it would have been reassuring to see the independent
confirmation. I give this one a B grade.

\vspace*{0.2cm}
\noindent
{\bf RJX1713.7-3946}
\vspace*{0.2cm}

The detection of TeV gamma-rays from this shell-type SNR was
reported by the CANGAROO group for the first time this year
\cite{muraishi99}. The observations were motivated by the
observation of a hard X-ray power-law spectrum by {\it ASCA}
\cite{koyama97}. In this respect, it is very similar to SN1006 but
is three times brighter in X-rays. It has a characteristic
dimension of 70 arc-min, lies at a distance of 1.1 kpc and has an
estimated age of 2,100 years. The $\gamma$-ray flux above 2 TeV is
$3 \times 10^{-12}$ photons cm$^{-2}$ s$^{-1}$ with a 5 $\sigma$
significance. There is evidence that the source is extended in the
same direction as the X-ray source. This is clearly a grade B source.

\vspace*{0.2cm}
\noindent
{\bf Cassiopeia A}
\vspace*{0.2cm}

It is natural that the strongest source in the radio sky should
have been one of the first targets of TeV observations
\cite{chudakov65}. It is appropriate that it should have been
eventually detected as a TeV source; however this only
came after a very long
exposure by the HEGRA group \cite{puhlhofer99}. As with SN1006
and RXJ1713.7-3946, these observations were motivated by
observations of a hard X-ray power-law spectrum \cite{allen97}. The
source is a classical shell-SNR of diameter 2.2 arc-min which is
effectively point-like to a $\gamma$-ray telescope. It is believed
to be 300 years old and there is no active pulsar at its center;
however there may be a neutron star. The HEGRA observations were
made in 1997 and 1998 and comprised some 130 hours on the source.
The flux above 1 TeV has not yet been determined but must be $\approx$ 
$3 \times 10^{-12}$ photons cm$^{-2}$ s$^{-1}$. The total detection
was just less than 5 $\sigma$ and it is probably the weakest TeV
source detected to date.

Upper limits to the TeV emission have been reported by the CAT
\cite{goret99} and Whipple \cite{lessard99} groups. These were at
lower energies but, because the exposures were much shorter, the
upper limits are compatible with the HEGRA detection. The three
results are summarized in Table \ref{casA}.

Because the detection is below the magic 5 $\sigma$ level and 
Cassiopeia A is the weakest
source yet detected (and hence more susceptible to systematic
effects), I give this a C grade for now.

\begin{table} 
\caption{Observations of Cassiopeia A}
\label{casA}

\begin{tabular}{cccl}
Group & {E$_{\rm th}$}
 & {Exposure} & {Integral Flux} \\ 
 & {(TeV))} & {(hours)} & 
{($10^{-11}$ photons cm$^{-2}$ s$^{-1}$)} \\
\tableline
Whipple & 500 & 7.5 & $<$ 0.66 \\
CAT     & 400 & 24.4 & $<$ 0.74 \\
HEGRA   & 1000 & 127.9 & $\approx$0.3 \\ 
\end{tabular}
\end{table}
 
\vspace*{0.2cm}
\noindent
{\bf Centaurus X-3} 
\vspace*{0.2cm}

New observations were reported on the high mass X-ray binary, Cen
X-3 \cite{chadwick99b}. The system contains a 4.8 s pulsar in orbit
with a period of 2.1 days. Originally reported as a source of
sporadic outbursts of pulsed emission
\cite{carraminano89,raubenheimer89}, it was later found to be a
source of steady (unpulsed) weak emission \cite{chadwick98}. At
this time it was also seen as an unpulsed GeV EGRET source
\cite{vestrand99}. The new observations, taken in 1998 and 1999 by
the Durham group, do not add to the overall statistical
significance of the detections which remain somewhat marginal;
hence I give it a C grade.

\subsection{Extragalactic Sources}
\vspace*{0.2cm}
\noindent
{\bf Markarian 421}
\vspace*{0.2cm}

Markarian 421 was one of the weakest AGNs detected by EGRET; it was
also the closest BL Lac at z = 0.031. It was the first TeV source
detected \cite{punch92}. It is also the AGN in which the clearest
correlations have been found over multiple wavelengths (see
\cite{cataneseweekes99}) and in which the shortest time variations
have been seen \cite{gaidos96}. At discovery, its intensity was
approximately 30\% of the Crab; however it has flared to levels
more than ten times greater than the Crab.

In 1998 there were extensive multiwavelength campaigns on this
source between various ground-based $\gamma$-ray observatories and the
{\it ASCA} and {\it BeppoSAX} X-ray satellites
\cite{takahashi99,maraschi99}. The most interesting event was the
flare seen on April 21, 1998 by the Whipple Observatory
\cite{catanese99} and  the {\it BeppoSAX} telescopes. Although
the flare was observed to rise and peak at the same time in both
telescopes, the TeV signal decayed within a few hours whereas the
X-ray signal persisted for half a day. It is difficult to model
this behavior.

The energy spectrum of Markarian 421 has been reported by several
groups. There is general agreement that it can be fit by a simple
power law. While the absolute flux has little meaning since it
varies with time, the differential power-law spectral index should
be comparable in different experiments unless it is also variable
with time. There is good agreement on the indices obtained thus far
by CAT ($-2.96 \pm 0.13 \pm 0.05$) \cite{piron99}; HEGRA ($-3.09
\pm 0.07 \pm 0.10$) \cite{aharonian99}; 7TA ($-2.81$)
\cite{7ta99a}. However the Whipple group gets consistently harder
spectra \cite{krennrich99} particularly during flaring e.g ($-2.54
\pm 0.04$) on May 7, 1996. Preliminary analysis of non-flaring data
gives a similar result. Obviously further work is required here to
ensure that the analysis is free of large systematic errors.

Despite its variability Markarian 421 is well-established and
merits an A grade.

\vspace*{0.2cm}
\noindent
{\bf Markarian 501}
\vspace*{0.2cm}

Markarian 501, a BL Lac at z=0.034, was the first extragalactic
$\gamma$-ray source detected first at TeV energies. Originally
detected as a weak source (8\% Crab) \cite{quinn95} it has been
intensively monitored by ground-based telescopes since then. The
TeV outburst from Markarian 501 in 1997 merited a Highlight session
at the 25th ICRC \cite{protheroe97}. Sadly while the conference was
taking place the source was already in decline and it has been
relatively quiescent ever since. Most of the interest
in the source since that time has been in a detailed analysis of
the high intensity signal, in particular in the derivation of an
accurate energy spectrum. 

The 1997 outburst data has been summarized in a number of
publications \cite{quinn99,aharonian9*,cat9*,7ta98}. Variations
with doubling times as short as two hours have been reported
\cite{quinn99} but in general the variations are not as short as
those seen in Markarian 421. There were no significant new results
from multiwavelength campaigns.

Spectral measurements were in agreement in that the energy spectrum
could not be satisfactorily represented by a simple power law.
The Whipple and HEGRA
groups \cite{hegra99,krennrich99}) reported that they observed no
change in spectral shape with source intensity; in contrast, the CAT
group \cite{tavernet99} using a simple Hardness ratio found that
the spectrum hardened as the intensity increased. It is not clear
whether this is a real change or the result of
the systematics in  a new analysis
method (which has not yet been tested on any other source). In addition
the CAT group reported that the weak intensity emission observed in
1998 could be best fit by a simple power law with differential
spectral index $ -2.97 \pm 0.20$.

Markarian 501 is the archetypical extreme BL Lac, characterized by
its low luminosity, its high synchrotron peak (up to 100 keV), and
its high Compton peak (up to TeV) \cite{catanese98,ghissellini99}.
As the strongest source (for a few months in 1997) thus far
observed, it clearly merits an A grade.

\vspace*{0.2cm}
\noindent
{\bf 1ES2344+514}
\vspace*{0.2cm}

Although less well-studied, this X-ray-selected BL Lac at z=0.044
is superficially very similar to the above two sources. Recent X-
ray observations by {\it Beppo-SAX} \cite{giommi99} emphasize this
similarity: time variability on times scales of hours has been seen
and the putative synchrotron spectrum peaks at energies greater
than 10 keV. It was reported as a TeV source \cite{catanese98}
primarily on the basis of a flare seen in one night at the 6
$\sigma$ level; the average flux over that night was F$_{\gamma}$
($>$350 GeV)
= $(6.6 \pm 1.9) \times 10^{-11}$ photons cm$^{-2}$ s$^{-1}$ which
was 60\% of the Crab. The averaged flux (including the flare) was
at the 5.8 $\sigma$ level. The source was not detected in the
1996/7 observing season.

Based on the observed behavior of Markarian 421 and Markarian 501
it might have been expected that continued monitoring of
1ES2344+514 would have confirmed this detection and given more
information about its properties at high energies. In practice,
continued monitoring by the Whipple group 
and HEGRA \cite{konopelko99b} have not confirmed
either the flaring or steady emission; hence this source which
would have been ranked B in 1997 must now have a C grade.

\vspace*{0.2cm}
\noindent
{\bf PKS2155-304}
\vspace*{0.2cm}

The above three sources are in the northern hemisphere; it had been
predicted that PKS2155-304 would be the best candidate for TeV
emission in the southern hemisphere. An X-ray-selected BL Lac, it
has been detected by EGRET and has been the object of
numerous multiwavelength observing campaigns. The Durham group
detected it in 1996 and 1997 \cite{chadwick99d}; the November 1997
observations were particularly interesting as they coincided with
observations by EGRET and RXTE which indicated that the source was
active at this time.

More recent observations by the Durham group (J.Osborne, this workshop) have
not detected the source. Because of its relatively large redshift
(z=0.116), the energy spectrum of this source is of particular
interest; however none is yet available. This appears to be a
highly probable source and thus merits a B grade.

\vspace*{0.2cm}
\noindent
{\bf 1ES1958-650}
\vspace*{0.2cm}

The Utah Seven Telescope Array group have reported the detection of
the BL Lac, 1ES1959+650 based on 57 hours of observation in 1998
\cite{7ta99a}. As with the four AGNs listed above, this is an X-
ray-selected BL Lac; its redshift is 0.048. The energy threshold
for these observations was 600 GeV. The flux level was not
reported but the total signal was at the 3.9 $\sigma$ level. This
is not normally considered high enough to claim the detection of a
new source; however, within this database there were two epochs
which were selected a posteriori  which gave signals above the
canonical 5 $\sigma$ level. This source has not yet been confirmed
by any other group; it was observed by the Whipple group but no
flux was detected \cite{95icrc}. It is therefore awarded
a B$-$ grade.

\vspace*{0.2cm}
\noindent
{\bf 3C66A}
\vspace*{0.2cm}

This is potentially the most exciting TeV detection of an AGN as it
is quite different from the other AGNs. The source is a radio-
selected, EGRET-detected, BL Lac and the redshift is 0.44, i.e.,
much more distant than the other objects. The Crimean Astrophysical
Observatory group using the GT-48 telescope detected this source at
the 5 $\sigma$ level in 1996 \cite{neshpor97}. The flux above 900
GeV was $(3 \pm 1) \times 10^{-11}$ photons cm$^{-2}$ s$^{-1}$.
There were previous and later upper limits to the TeV emission from
the source, e.g. F$_{\gamma}$ ($>$ 350 GeV) $<$ $1.9 \times 10^{-11}$ photons-
cm$^{-2}$-s$^{-1}$ from Whipple in 1993 \cite{kerrick95}. Confirmation
of this detection is urgently required; until then it must be
considered a grade C source.

\section{Periodicity in 1997 signal from Mrk 501}

Several groups have reported on the apparent periodicity in the TeV
$\gamma$-ray signal from Markarian 501 (TA, HEGRA, Whipple). The
best data base is that of the HEGRA group since they observed
during part of the bright period of the moon with one of their
telescopes and hence have a database that is less prone to aliases.
The reported periodicities occured at 12.7 day \cite{7ta98} and 23-24
day \cite{7ta99b,hegra99a} and S.Fegan, this workshop,
 and were arrived at using the
Lomb method which is recommended for observations made at irregular
intervals. The epoch chosen by the HEGRA group for periodicity
analysis is {\it a posteriori} but coincides with the bulk of the TeV
observations and the peak in the $\gamma$-ray signal intensity.
There is no evidence for periodicity outside this interval, either
in 1997 or in other years. A visual inspection shows that the
$\gamma$-ray signal has a few clearly defined flares with several
time constants and the most obvious is at 23 days. 

Since all the $\gamma$-ray experiments were observing at
approximately the same time, they must see the same time
variations; hence reports from the separate experiments do not
constitute independent confirmations. The real question is whether
the observed "periodicity" is really statistically significant
given the large number of time variations. It is difficult to
arrive at the real statistical significance of the observed effect.

Similar periodicity is seen in the X-ray signal by ASM/ RXTE
and it has been suggested that this constitutes independent
evidence for the periodicity. However, correlation between the X-ray
and TeV $\gamma$-ray signals from Markarian 421 and 501 on a
variety of time-scales now seems to be well-established so that the
independent analysis of the RXTE database only confirms this
correlation, not the statistical significance of the periodicity.

The conclusion is that while there is apparent periodicity in the
TeV/X-ray signals from Markarian 501 for a five month epoch in
1997, it is almost impossible to arrive at a 
definitive conclusion about its
statistical significance. 

Those who survived the many pseudo-periodicities seen in the "a
posteriori" analysis of the TeV observation of binaries in the
previous decade will perhaps be forgiven a little skepticism in the
discussion of this new and potentially important result. It is
unlikely that it will become credible until the phenomenon is
observed again, either in Markarian 501 or another BL Lac.

\section{Gamma Ray Burst GRB970417a}

At this workshop there was the first report of the detection of a
$\gamma$-ray burst at TeV energies by Milagrito, the first stage 
of the Milagro experiment (J.McEnery, this workshop). In some 15
months of operation 54 possible GRBs were within the FOV of the
detector and from one of these , GRB970417a, 18 events were
detected during the duration of the BATSE burst (9 s) where the
background was 3.46 events. Allowing for trials, the probability of
the observation being a statistical fluctuation was calculated to
be $1.5 \times 10^{-3}$. The energy threshold was about 1 TeV and
the flux a few times $10^{-12}$ photons- cm$^{-2}$-s$^{-1}$. Given the
importance of a detection of a TeV burst it would seem wise to
await a confirmation from the full, more sensitive, Milagro array
before too many conclusions are drawn.  The BATSE fluence of
GRB970417a was $1.5 \times 10^{-7}$ erg cm$^{-2}$, a rather weak burst, and
there was nothing otherwise unusual about it. There was no precise
position and hence no information on X-ray or optical counterparts
nor any indication as to distance.

\begin{table}
\caption{The TeV Source Catalog c.1999}
\label{catalog}
\begin{center}
\begin{tabular}{lllll} \hline  

Source &   Type &    Discovery & EGRET & Grade  \\

Galactic &               &              &         &          \\

Crab Nebula &       Plerion &      1989 &      yes &   A  \\

PSR 1706-44 &       Plerion? &     1995 &  no &    A
\\

Vela &              Plerion? &     1997 &  no &    B
\\

SN1006 &            Shell &        1997 &  no &   B$-$ \\

RXJ1713.7-3946 &    Shell &        1999 &  no &   B \\

Casssiopea A   &    Shell &        1999 &  no &   C \\

Centuarus X-3  &    Binary &       1999 &  yes &  C \\

Extragalactic &          &              &    &       \\

Markarian 421  &    XBL z=0.031 &  1992 &  yes &  A \\

Markarian 501  &    XBL  z=0.034 & 1995 &  yes &  A \\

1ES2344+514    &    XBL  z=0.044 & 1997 & no  &   C \\

PKS2155-304    &    XBL  z=0.116 & 1999 & yes &  B \\

PKS1959+650    &    XBL  z=0.048 & 1999 &  no &   B$-$ \\

3C66A          &    RBL  z=0.44 &  1998 &  yes & C \\
\end{tabular}
\end{center}
\end{table}

\section{HE/VHE status and outlook}

Based on the above discussion the 1999 TeV Source catalog is
derived (Table \ref{catalog}); it is disappointing that it is not
much larger than the 1997 Catalog.

As we conclude this century it is worthwhile to summarize the
progress in HE and VHE astronomy and compare the achievements in
each band. The recent publication of the 3rd EGRET Catalog
summarizes the field at energies from 30 MeV to 10 GeV. 

In Table \ref{status} the number of sources reported in various
categories is compared. The 100 MeV sources are from the 
Third EGRET Catalog \cite{hartmann99} and from 
\cite{esposito9*}. Don Kniffen has pointed out (this
workshop) that only a handful of the EGRET sources could be
classified with an A grade.
The TeV sources are from Table \ref{catalog}; note in this
context C must be considered a passing grade.

\begin{table}
\caption{Status of HE/VHE Sources}
\label{status}

\begin{tabular}{lll} \hline  

Energy Range   &         10 MeV - 10 GeV &   300 GeV - 30 TeV \\
\hline
Platform       &              Space      &             Ground \\
\hline

Discrete Sources &  & \\

Type           &  No. of Sources & No. of
Sources \\
\hline

AGNs           &         75        &              6     \\

Normal Galaxies     &         1    &              0     \\

Pulsars   &              5?   &                   0     \\

SNR Shell      &         4?   &                   3    \\

SNR Plerion    &         1    &                   3    \\

Binaries       &         1    &                   1    \\

Total identified    &    87   &                   13   \\

Unidentified        &    165  &                   0    \\

Total          &         250  &                   13   \\
\hline
Other Sources  &              &                        \\

Galactic Plane &         Yes  &                   No   \\

Extragalactic Diffuse    &         Yes  &    No        \\

All Sky Survey &         Yes  &                   No   \\

Gamma Ray Bursts    &    5    &                   1?   \\

\end{tabular}

\end{table}

Although EGRET has still some sensitivity the mission is
essentially over and not much change can be expected in the
observational picture until the launch of GLAST in 2005
(hopefully). The intermediate missions, AMS and AGILE, which are
described elsewhere in these proceedings, will not significantly
improve on the EGRET flux sensitivity and can be considered place-
holders for GLAST. The latter will offer an improvement of a factor
of 10-20 in most parameters compared to EGRET. 

In contrast to the drought expected in MeV-GeV $\gamma$-ray
observations in the immediate future, ground-based $\gamma$-ray
astronomy has never been more active. There are already nine
atmospheric Cherenkov imaging telescopes in operation and two 
low threshold air
shower arrays; one can expect to see steady improvements in
sensitivity in these telescopes over the next decade. In addition,
there are several other Cherenkov experiments coming on-line e.g.,
STACEE, CELESTE, Solar-Two, Pachmari, etc. Four major Cherenkov
imaging telescopes/arrays are scheduled for completion by 2003,
well in advance of GLAST and with significant overlap  in
the 30-300 GeV range. One can expect to see a steady increase in
the GeV-TeV source catalog from ground-based observations so that
even if the GLAST launch were to be delayed there would be a
healthy increase in activity in studies of $\gamma$-ray
astrophysics at these high energies. 
\begin{table}
\caption{Future Roadmap for HE/VHE Gamma Ray Astronomy}
\begin{center}
\begin{tabular}{lllllll}

Energy    &         &         &         &         &         & \\
       &    MeV  &    GeV  &    GeV  &    GeV  & TeV &  TeV \\
       & 10-100  &   0.1-1    &  1-10   & 10-100  & 0.1-1  & 1-10+  \\
       & Space  & Space   & Space    & Space/  &    &    \\   
          &         &         &   & Ground & Ground & 
Ground    \\
\hline
Year & & & &                     & & \\

1999 &*Comptel* & (EGRET) &  &********** & **9ACITs* &***+2ASA \\

2000 &****     &    &    & CEL/STAC &********** &********** \\

2001 &****     &    &    & **********&********** &********** \\

2002 &Integral &    &    & **********&**********&**********\\

2003 &** &AMS/AGILE&**********&*MAGIC**&**********&********** \\

2003 &**  &***********&**********&**********&HESS/CAN &**********\\

2004 &**  &***********&**********&**********&VERITAS* &********** \\

2005 &~*GLAST**&**GLAST**& **GLAST*&*GLAST**&**********&********** \\

2006 &~*********&***********&**********&**********&**********&********** \\

2007 &~*********&***********&**********&**********&**********&********** \\

2008 &~*********&***********&**********&**********&**********&********** \\
\end{tabular}
\end{center}
\end{table}

\section{Where have all the hadrons gone?}

It is a matter of some disappointment for the many cosmic-ray
physicists who entered the field of high energy $\gamma$-ray
astronomy that none of the sources thus far detected, either at HE
or VHE energies, can be positively identified with hadron
progenitors. In the early days it was widely believed that
$\gamma$-ray astronomy would finally solve the mystery of the
origin of the cosmic radiation. However with the exception of the
galactic plane (and the Large Magellanic Cloud) where we observe
not the source of cosmic radiation but its 
interaction during propagation, every one
of the sources detected so far can be attributed to a source in
which electrons are the progenitor particles. In no source is the
much heralded "bump" in the energy spectrum near 70 MeV seen. In
some cases there are proponents of plausible models in which
hadrons are the progenitors but there are equally vociferous
proponents who would advocate electron models and in many cases
these seem the more plausible. Thus in the 40 plus years since the
publication of Morrison's seminal paper  \cite{morrison58} while we
have learnt some interesting astrophysics we have come no closer to
a definitive model of cosmic-ray origins.

\section{Acknowledgements}

Research in VHE $\gamma$-ray astronomy at the Smithsonian
Astrophysical Observatory is supported by a grant from the U.S.
Department of Energy. Mike Catanese and Vladimir Vassiliev 
read the manuscript and
supplied many critical comments.

\end{document}